\documentclass[superscriptaddress,aip,apl,reprint,floatfix,citeautoscript,showpacs,floatfix]{revtex4-1}

\usepackage{graphicx}
\usepackage{epsfig}
\usepackage{color}
\usepackage{latexsym}
\usepackage{amsmath,bm}
\usepackage{natbib}
\usepackage{rotating}
\usepackage{multirow}
\usepackage{array}

\newcommand{\etal}{\textit{et al.}}

\newcommand{\CISSe}{CuIn(S,Se)$_2$}

\newcommand{\CZTS}{Cu$_2$ZnSnS$_4$}
\newcommand{\CZTSe}{Cu$_2$ZnSnSe$_4$}
\newcommand{\CZTSSe}{Cu$_2$ZnSn(S,Se)$_4$}
\newcommand{\abinitio}{\textit{ab initio}}
\newcommand{\GW}{\ensuremath{GW}}

\newcommand{\scGW}{sc\GW}
\newcommand{\lsi}{Laboratoire des Solides Irradi\'es and ETSF, \'Ecole Polytechnique,
CNRS, CEA-DSM, 91128 Palaiseau, France}
\newcommand{\lpmcn}{LPMCN and ETSF, Universit\'e Claude Bernard Lyon I and CNRS, 69622
Villeurbanne, France}

\begin{document}

\title{Band structures of Cu$_2$ZnSnS$_4$ and Cu$_2$ZnSnSe$_4$ from many-body methods}

\author{Silvana Botti}
\affiliation{\lsi}
\affiliation{\lpmcn}

\author{David Kammerlander}
\affiliation{\lpmcn}

\author{Miguel A. L. Marques}
\affiliation{\lpmcn}

\date{\today}

\begin{abstract}
We calculate the band structures of kesterite and stannite
Cu$_2$ZnSnS$_4$ and Cu$_2$ZnSnSe$_4$, using a state-of-the-art self-consistent \GW\
approach. Our accurate quasiparticle states allow to discuss: the
dependence of the gap on the anion displacement; the key-role of the
non-locality of the exchange-correlation potential to obtain good
structural parameters; the reliability of less expensive hybrid
functional and GGA+U approaches. In particular, we show that even if
the band gap is correctly reproduced by hybrid functionals, the
band-edge corrections are in disagreement with self-consistent GW
results, which has decisive implications for the positioning of the
defect levels in the band gap.
\end{abstract}


\maketitle

Thin-film solar cells made of Cu(In,Ga)Se$_2$ (CIGS) ally cost
reduction and high efficiency, and compete today as successors of the
dominating silicon technology. Nevertheless, there are concerns about
their large scale production due to the increasing price of In and
Ga. Quaternary chalcogenides \CZTSSe\ (CZTS) have recently been
proposed as alternative absorbers. Their crystal structures and
electronic properties are very similar to those of the parent CIGS,
while their constituent elements are naturally abundant and
non-toxic. The alloys Cu$_2$ZnSnS$_x$Se$_{1-x}$ have
optimal gaps according to the Shockley-Queisser
limit and their use as absorbers in
thin film solar cells is getting established by a
growing energy conversion efficiency (almost
10\%~\cite{todorov10,redinger11} for lab cells). However, the understanding
of the properties of the different phases of \CZTS\ and \CZTSe
is still rather superficial, and only few recent studies have addressed
their structural~\cite{persson10,chen09,chen10bis},
electronic~\cite{chen09,persson10,paier09}, and defect
properties~\cite{chen10,nagoya10}.

The zincblende-derived kesterite structure ($I\bar 4$) of CZTS is
recognized to be the most stable. The energy difference per atom with
respect to the stannite structure ($I\bar 4 2m$) is only of few meV
per atom~\cite{persson10,chen09,chen09bis}, proving that kesterite and
stannite phases can coexist in experimental samples, and explaining
the reported disordered structures~\cite{schorr07}.  Similarly to
chalcopyrites CIGS, quaternary kesterite and stannite CZTS are
obtained from the zincblende structure by replacing the Zn cations in
such a way that each anion (Se or S) is coordinated by one Zn, one Sn
and two Cu atoms. The existence of three distinct cations results in
three different cation-anion bond lengths, which induce a displacement
of the anion from its ideal zincblende site. That distortion is
measured by the anion displacement parameters ($u_x$, $u_y$, $u_z$),
i.e., the relative coordinates of the anion in the conventional
body-centered tetragonal cell. The anion displacement is harder to
measure than the lattice constants $a$ and $c$, due to the
inhomogeneity of the samples. Indeed, the dispersion of data for $u$
is known to be large in CIGS compounds~\cite{jaffe84,merino96}.
Concerning CZTS, fewer measurements can be found in literature and in
most cases only $a$ and $c$ are reported~\cite{exp-geo}.
Concerning the experimental band
gaps~\cite{kes-S-gaps,stan-S-gaps,ahn10}, early studies suggested
a value of about 1.4--1.6\,eV both for S and Se compounds. This was in
disagreement with density functional theory (DFT)
calculations~\cite{chen09,persson10}, using semi-local or hybrid functionals, 
which obtained systematically a
smaller gap for Se compounds. More recent measurements~\cite{ahn10} delineate
a gap of about 1\,eV for \CZTSe, explaining the previous
overestimation with the presence of ZnSe in the sample.

Note that the current workhorse \abinitio\ theory is, in fact, DFT, in
combination with the local density approximation (LDA) or generalized
gradient approximations (GGA). However, this approach is totally
inadequate to study the electronic structure of materials where the
band gap is controlled by the hybridization of the $d$ states of a
transition metal with $p$ states, such as in CIGS or \CZTS. Moreover
the anion-cation bonds can be poorly described, leading to anion
displacements outside the experimental range, even if lattice
constants are as usual well reproduced.  In particular, for \CISSe,
LDA and GGA yield structural internal parameters with an error of
about 5\% for $u$~\cite{jaffe84,merino96,vidal10}. Such an error leads
to an underestimate by more than 50\% of the band gap in a calculation
at the theoretical geometry~\cite{vidal10}.

\begin{figure}[t]
 \centering \includegraphics[clip,width=0.9\columnwidth]{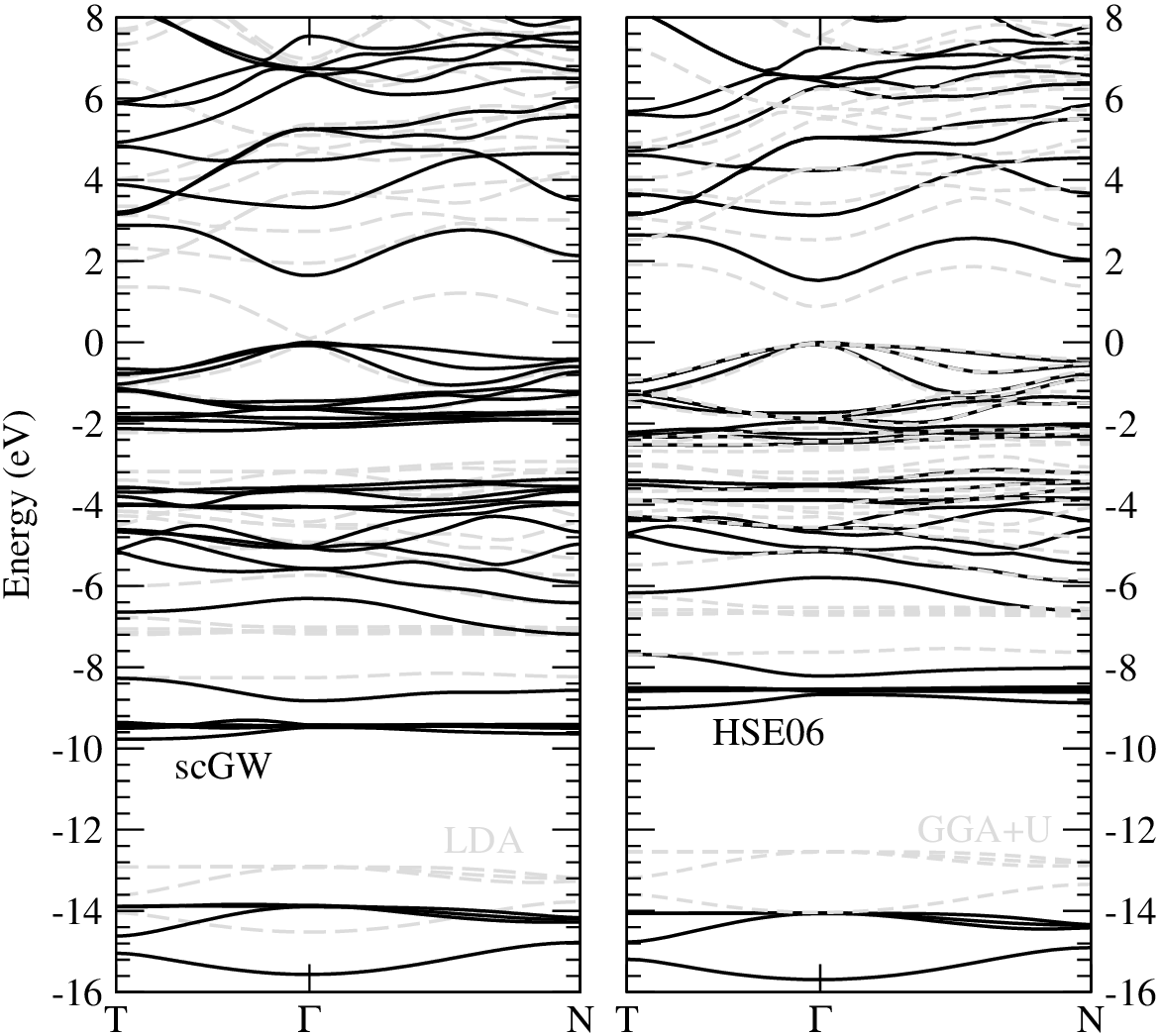}
 \caption{Band structures for kesterite
   \CZTS\ calculated using: (left panel) DFT-LDA (dashed line)
   and \scGW\ (continuous line), (right panel) GGA+$U$ (dashed
   line) and HSE06 (continuous line). Top valence bands are set to zero.}
 \label{fig:bands-compare}
\end{figure}

In this letter, we solve these problems by using state-of-the-art
\abinitio\ approaches that go beyond standard DFT to determine
accurate quasiparticle band structures of both kesterite and stannite
\CZTS\ and \CZTSe.  These calculations are based on a restricted
self-consistent (sc) \GW\ scheme, which has the advantage of being
independent of the starting point (i.e., the poor LDA Kohn-Sham
states) at the price of a larger computational complexity. Such
approach, that we will refer to as \scGW, consists in performing a
self-consistent \GW calculation within the Coulomb hole plus screened
exchange (COHSEX) approximation~\cite{hedin}, followed by a
perturbative \GW\ on top of it. This method has been applied to many
transition-metal compounds, yielding excellent results for the band
gaps and the quasiparticle band
structure~\cite{bruneval,vidal10,vidal10bis}.  Standard LDA
or GGA, and \scGW\ calculations were performed using the code
ABINIT. We included semicore states in the valence to
build the norm-conserving pseudopotentials for Cu, Zn and Sn. We also
used the code VASP~\cite{vasp} for Heyd-Scuseria-Ernzerhof
(HSE06)\cite{HSE06} hybrid functional and GGA+U calculations. Due to
the similarity of CZTS materials with the CIGS family, the convergence
parameters turned out to be the same reported in
Refs.~\onlinecite{vidal10}.

In Fig.~\ref{fig:bands-compare} we display band structures for
kesterite \CZTS, obtained using different theoretical schemes at the
experimental geometry~\cite{exp-geo}.  In the left panel, the
Kohn-Sham LDA band structure is compared with the
\scGW\ bands. We observe that \scGW\ corrections upshift
almost rigidly the lowest conduction states.  Concerning the valence:
(i) the dispersion of the S $p$--Cu $d$ antibonding states at the top of
the valence remains fairly unaltered, even if the overlap of LDA and
quasiparticle wavefunctions shows remarkable variations in this
region. (ii) The band width of the S $p$--Cu $d$ bonding states (located
between -3.5 and -6.7\,eV) slightly increases. (iii) The bands
associated to the (Zn,Sn)-S bond (between -8 and -10\,eV) are inverted
and downshifted by about 2\,eV with respect to LDA. (iv) Also the S$s$
states are moved down by about 2\,eV.  In the right panel of
Fig.~\ref{fig:bands-compare} we show the same bands as obtained from
GGA+U and HSE06 calculations. In this case HSE06 bands are remarkably
similar to \scGW\ bands. This is not particularly surprising as the
Hartree-Fock mixing of HSE06 is particularly suited for materials with
gaps of about 1--2\,eV~\cite{paier08,marques11}.  As expected, GGA+U
shifts down the states with Cu $d$ character, thereby opening the gap
to a reasonable value. However, it is evident from the figure, that
the overall description of the band dispersions is quite inaccurate.

\begin{figure}[t]
 \centering \includegraphics[clip,width=0.9\columnwidth]{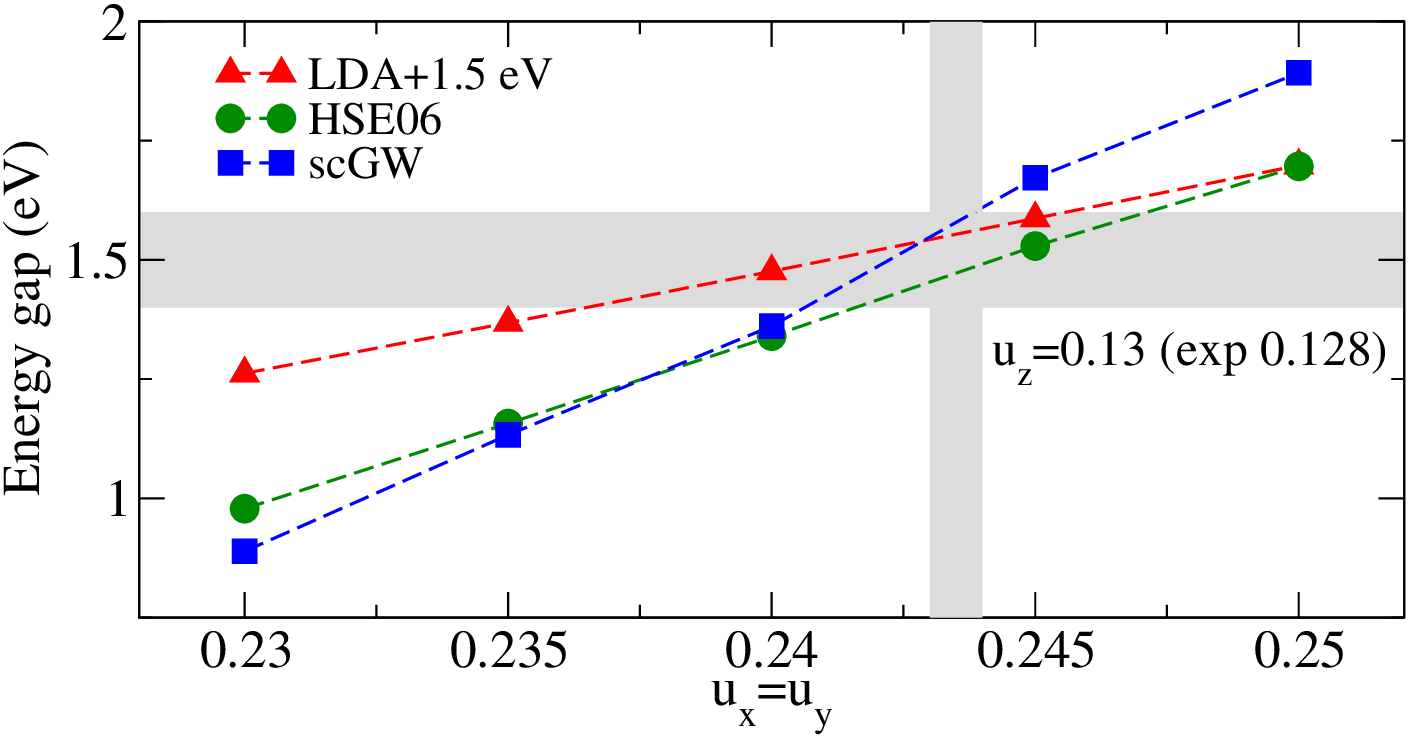}
 \caption{ (Color online) Quasiparticle band gap vs. anion
   displacement $u_x=u_y$ for kesterite \CZTS, using DFT-LDA (red
   triangles), HSE06 (green circles) and \scGW\ (blue
   squares). DFT-LDA values are upshifted by 1.5 \,eV. The vertical
   (horizontal) shaded areas give the spread of experimental data for
   $u$ (band gap).}
 \label{fig:gap-vs-u}
\end{figure}

It is by now known that the band gap in CIGS materials is extremely
sensitive to structural distortions~\cite{jaffe84} and the remarkable
stability of the band gap found experimentally can only be explained
by compensating effects induced by intrinsic defects~\cite{vidal10}.
In order to establish if a similar behavior is also found in CZTS
compounds, we performed calculations for kesterite (see
Fig.~\ref{fig:gap-vs-u}) and stannite \CZTS\ by varying the anion
displacements. We also verified that sensible variations of $a$ and $c$
lattice parameters have negligible effects on the gap, in analogy to
the case of CIGS~\cite{vidal10}. We observe a strong variation of the band gap with
$u$, which is due to similar variations of both valence and conduction
band edges.  Note that the position of the top valence has important
implications for the formation energies of charged defects.  The
slopes are substantially larger for \scGW\ calculations than for
DFT-LDA, and even than for HSE06. In fact, they are controlled by the
screening, which is essential to include in a self-consistent way, as
in \scGW.

\begin{figure}[t]
 \centering
 \includegraphics[clip,width=0.9\columnwidth]{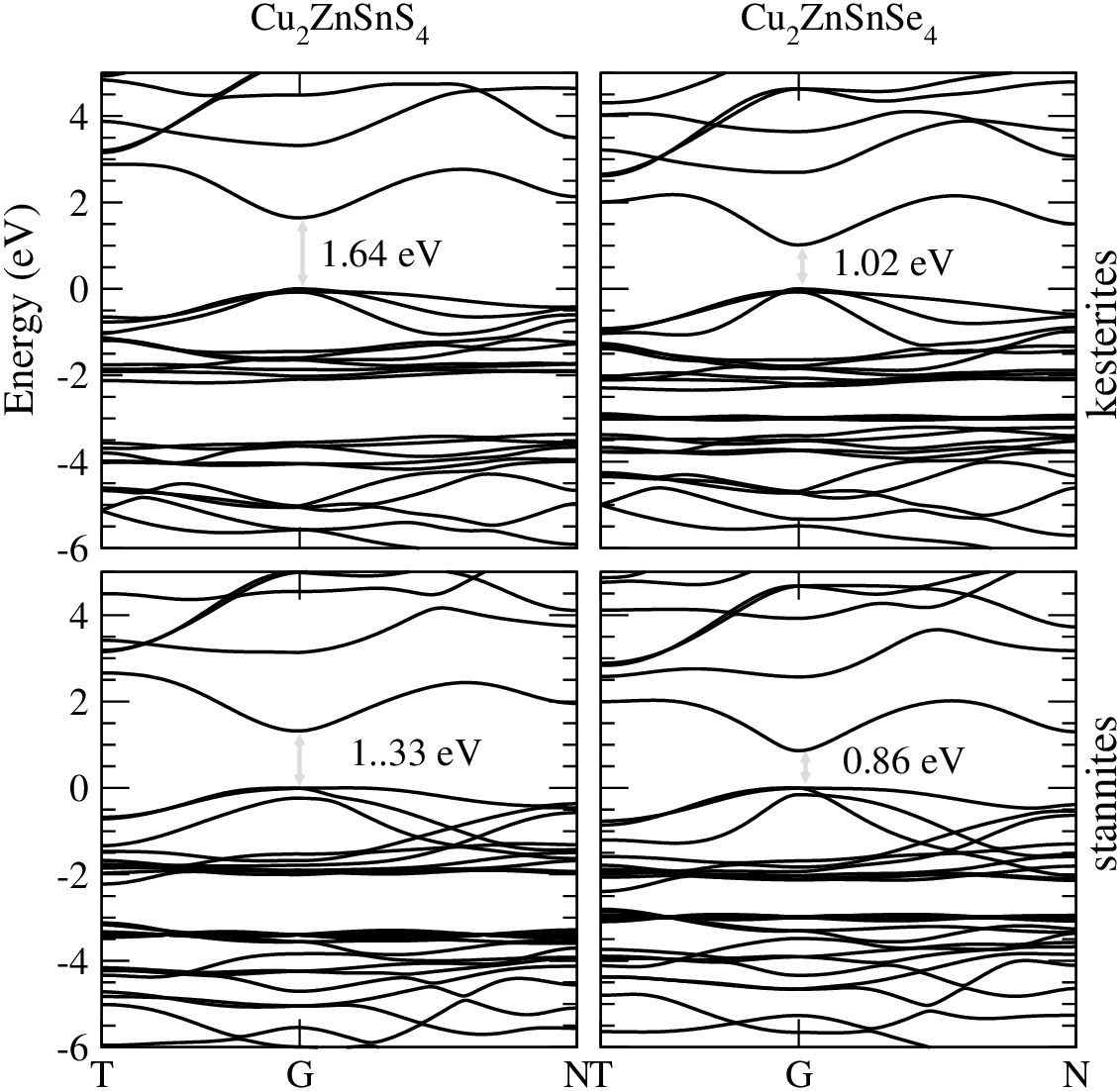}
 \caption{Band structures from \scGW\ for (a) kesterite \CZTS, (b) stannite \CZTS, (c)
   kesterite \CZTSe, (d) stannite \CZTSe.}
 \label{fig:bands-scGW}
\end{figure}

In Fig.~\ref{fig:bands-scGW} we can see the \scGW\ band structures of
the four compounds of the CZTS family. For the kesterite \CZTS\ and
stannite \CZTSe\ the experimental geometries were
used~\cite{exp-geo}. For the remaining compounds, the experimental
anion displacements are not reported~\cite{exp-geo}, and it was
therefore necessary to resort to the theoretical structures. In view
of the strong variation of the gap mentioned before, the choice of the
theoretical framework for the geometry optimization is essential. In
fact, we verified that LDA/GGA relaxed structures lead to unacceptably large errors
(up to 40\%) in the sc\GW\ band gap, essentially related to the error in
$u$: Cu--(S,Se) and Zn--(S,Se) bond lengths are too small in DFT-LDA, while Sn--(S,Se) 
bond lengths are too large. The solution to this issue relies on the use of the HSE06
functional, that yields extremely accurate values for $u$ and the cell
parameters, thanks to a better description of the localized states participating in the bonds.
The major difference among the four band structures of
Fig.~\ref{fig:bands-scGW} is the width of the band gaps, which are in
excellent agreement with experimental results. As in the CIGS family,
the Se compounds have a smaller gap with respect to the S ones.
Furthermore, stannites have consistently smaller gaps than kesterites.

\begin{table}
\caption{Band gaps and valence-band shifts with respect to LDA
(in eV) for all the structures considered in this work. \label{tab:band_edges}}
\begin{tabular}{m{0.2cm}m{0.2cm}r|rrrrr}
&&                    & LDA  & GGA+U & HSE & \scGW & exp. \\
  \hline
\multirow{4}{*}{\begin{sideways} kesterite \end{sideways}} &
\multirow{3}{*}{\begin{sideways} ~~S \end{sideways}}
&  $E_{\rm g}$        & 0.09   &  0.86 &  1.52 &  1.64 & 1.4--1.6~\cite{kes-S-gaps} \\
&&  $\Delta E_{\rm v}$ & 0.00   & -0.52 & -0.81 & -0.49 & \\   
\cline{3-8}
&\multirow{3}{*}{\begin{sideways} ~~Se \end{sideways}}
&  $E_{\rm g}$        & -0.24    & 0.34  & 0.94 &  1.02  & 0.8--1.0~\cite{ahn10}  \\
&&  $\Delta E_{\rm v}$ & 0.00   & -0.27 & -0.55 &  -0.31& \\   
\hline
\multirow{4}{*}{\begin{sideways} stannite \end{sideways}} &
\multirow{3}{*}{\begin{sideways} ~~S \end{sideways}}
&  $E_{\rm g}$      &  -0.11  & 0.63 & 1.27  &    1.33 & 1.4--1.5~\cite{stan-S-gaps}  \\
&&  $\Delta E_{\rm v}$ & 0.00  & -0.53 & -0.80  &   -0.42 & \\   
\cline{3-8}
&\multirow{3}{*}{\begin{sideways} ~~Se \end{sideways}}
&  $E_{\rm g}$        & -0.41 & 0.15  & 0.75   &  0.87  & 0.8--1.0~\cite{ahn10} \\
&&  $\Delta E_{\rm v}$ & 0.0   & -0.28 & -0.56  &  -0.35 &  \\   
\end{tabular}
\end{table}

Finally, in table~\ref{tab:band_edges} we show the band gaps and the
top valence shifts $\Delta E_{\rm v}$ with respect to DFT-LDA,
obtained by aligning the average electrostatic potential in the
different theoretical schemes. The validity of this alignment 
scheme is discussed in Ref.~\onlinecite{pasquarello}.  Band edge-shifts are
essential quantities to determine the position of defect levels in the
gap and band offsets at interfaces. The most striking feature is that,
even if HSE06 gives very good band gaps, the top valence and bottom
conduction bands are systematically too low.  These results suggest
that the fact that hybrids (and more in general tuned hybrids) can
give good gaps, does not mean that they always are able to reproduce
correctly band-edge shifts~\cite{pasquarello}. We observe also that
the valence band shift in GGA+U is close to the one obtained using
\scGW, despite the underestimation of the gap.  Note however, that
GGA+U does not account for the dependence on $u$ of the valence-band
shift, that is particularly strong in this family.

In conclusion, we obtained from many-body calculations a very accurate
description of the electronic properties of the CZTS family. We proved
that the HSE06 hybrid functional offers a good compromise between
accuracy and computational cost, yielding good gaps and relaxed
structures. Nevertheless, it cannot assure a reliable description of
the valence and conduction contributions to the band-edge corrections,
which has decisive implications for the position of the defect levels
in the corrected band gap.

Furthermore we found a remarkable similarity between the electronic
properties of the stannites and kesterite compounds, and to a large
extent between these and the CIGS family. 
This once more points to the direction that CZTS materials are excellent
candidates to replace the more costly CIGS in the absorbing layer of
thin film photovoltaic cells.

SB acknowledges support from EUs 7th Framework Programme (e-I3
contract ETSF), and MALM from the French ANR
(ANR-08-CEXC8-008-01). Calculations were performed at GENCI (project
x2010096017).


\begin{thebibliography}{ll}



\bibitem{todorov10} 
K. Todorov, K.\,B. Reuter, and D.\,B. Mitzi,
  Adv. Mater. {\bf 22}, E156 (2010).

\bibitem{redinger11} 
A. Redinger, D.\,M. Berg, P.\,J. Dale, and S. Siebentritt, 
  J. Am. Chem. Soc. {\bf 133}, 3320 (2011).

\bibitem{persson10} 
C. Persson, 
  J. of Appl. Phys. {\bf 107}, 053710 (2010).

\bibitem{chen09} 
S. Chen, X.\,G. Gong, A. Walsh, and S.-H. Wei, 
  Appl. Phys. Lett. {\bf 94}, 041903 (2009).

\bibitem{chen10bis} 
S. Chen, A. Walsh, Y. Luo, J.-H. Yang, X.G. Gong, and S.-H. Wei,
  Phys. Rev. B {\bf 82},  195203 (2010).

\bibitem{paier09} 
J. Paier, R. Asahi, R. Wahl, and  G. Kresse, 
  Phys. Rev. B {\bf 79}, 115126 (2009).

\bibitem{chen10} 
S. Chen, X.\,G. Gong, A. Walsh, and S.-H. Wei, 
  Appl. Phys. Lett. {\bf 96} 021902 (2010); 
S. Chen, J.-H. Yang, X.\,G. Gong, A. Walsh, and S.-H. Wei, 
  Phys. Rev. B {\bf 81}, 245204 (2010).

\bibitem{nagoya10} 
A. Nagoya, R. Asahi, R. Wahl, and G. Kresse, 
  Phys. Rev. B {\bf 81}, 113202 (2010).

\bibitem{chen09bis} 
S. Chen, X.\,G. Gong, A. Walsh, and S.-H. Wei, 
  Phys. Rev. B {\bf 79}, 165211 (2009).

\bibitem{schorr07} 
S. Schorr, 
  Thin Solid Films {\bf 515} (2007).

\bibitem{jaffe84} 
J.\,E. Jaffe and A. Zunger, 
  Phys. Rev. B {\bf29}, 1882 (1984).

\bibitem{merino96} 
J.\,M. Merino, J.\,M. de Vidales, S. Mahanty, R. D\'\i az, F. Rueda, and M. L\'eon, 
  J. Appl. Phys. {\bf 80}, 5610 (1996).

\bibitem{exp-geo} 
S.\,R. Hall, J.\,T. Szymanski, and J.\,M. Stewart, 
  Can. Mineral. {\bf 16}, 131 (1978); 
H. Hahn and H. Schulze, 
  Naturwiss. {\bf 52}, 426 (1965); 
G.\,S. Babu, Y.\,B.\,K. Kumar, P.\,U. Bhaskar, and V.\,S. Raja, 
  Semicond. Sci. Technol. {\bf 23}, 085023 (2008); 
D. Olekseyuk, L.\,D. Gulay, I.\,V. Dydchak, L.\,V. Piskach, O.\,V. Parasyuk, and O.\,V. Marchuk, 
  J. Alloys Compd. {\bf 340}, 141 (2002).

\bibitem{kes-S-gaps} 
H. Katagiri, K. Saitoh, T. Washio, H. Shinohara, T. Kurumadani, and S. Miyajima, 
  Sol. Energy Mater. Sol. Cells {\bf 65}, 141 (2001);
J.-S. Seol, S.-Y. Lee, J.-C. Lee, H.-D. Nam, and K.-H. Kim, 
  Sol. Energy Mater. Sol. Cells {\bf 75}, 155 (2003); 
T. Tanaka, T. Nagatomo, D. Kawasaki, M. Nishio, Q. Guo, A. Wakahara, A. Yoshida, and H. Ogawa, 
  J. Phys. Chem. Solids {\bf 66}, 1978 (2005);
N. Kamoun, H. Bouzouita, and B. Rezig, 
  Thin Solid Films {\bf 515}, 5949 (2007); 

\bibitem{stan-S-gaps} 
N. Nakayama and K. Ito, 
  Appl. Surf. Sci. 92, {\bf 171} (1996); 
J. Zhang, L. Shao, Y. Fu, and E. Xie, 
  Rare Metals {\bf 25}, 315 (2006).

\bibitem{ahn10} 
  S. Ahn \etal,
  Appl. Phys. Lett. {\bf 97}, 021905 (2010) and references therein.

\bibitem{vidal10} 
J. Vidal, S. Botti, P. Olsson, J.-F. Guillemoles, and L. Reining, 
  Phys. Rev. Lett. {\bf 104}, 056401 (2010).


\bibitem{hedin} 
L. Hedin and S. Lundqvist, 
  Solid State Phys. 23, {\bf 1} (1970).

\bibitem{bruneval} 
F. Bruneval, N. Vast, and L. Reining, 
  Phys. Rev. B {\it 74}, 045102 (2006); 

\bibitem{vidal10bis} 
J. Vidal, F. Trani, F. Bruneval, M. A. L. Marques, and S. Botti,
  Phys. Rev. Lett. {\bf 104}, 136401 (2010).


\bibitem{vasp} 
G. Kresse and J. Furthm\"uller, 
  Comput. Mater. Sci. {\bf 6}, 15 (1996). 

\bibitem{HSE06} 
J. Heyd, G.\,E. Scuseria, and M. Ernzerhof, 
  J. Chem. Phys. {\bf 118}, 8207 (2003); {\bf 124}, 219906(E) (2006). 

\bibitem{paier08} 
J. Paier, M. Marsman and G. Kresse, 
  Phys. Rev. B {\bf 78}, 121201 (2008).

\bibitem{marques11} 
M.\,A.\,L. Marques, J. Vidal, M.\,J.\,T. Oliveira, L. Reining, S. Botti,
Phys. Rev. B {\bf 83}, 035119 (2011).

\bibitem{pasquarello}
A. Alkauskas, P. Broqvist, and A. Pasquarello,
Phys. Status Solidi B, {\bf 248}, 775 (2011).

\end{thebibliography}
\end{document}